\documentclass[a4paper]{jpconf}
\usepackage{graphicx}

\newcommand{\ldm}{\Delta m_{31}^2}

\newcommand{\vldm}{\Delta m_{41}^2}

\begin{document}
\title{Phenomenology of neutrino oscillations at the neutrino factory}

\author{Jian Tang}

\address{Institut f{\"u}r Theoretische Physik und Astrophysik, 
       Universit{\"a}t W{\"u}rzburg, \\
       D-97074 W{\"u}rzburg, Germany}

\ead{jtang@physik.uni-wuerzburg.de}

\begin{abstract}
We consider the prospects for a neutrino factory to measure mixing angles, the CP violating phase and mass-squared differences by detecting wrong-charge muons arising from the chain $\mu^+\to\nu_e\to\nu_\mu\to\mu^-$ and the right-charge muons coming from the chain $\mu^+\to\bar{\nu}_\mu\to\bar{\nu}_\mu\to\mu^+$ (similar to $\mu^-$ chains), where $\nu_e\to\nu_\mu$ and $\bar{\nu}_\mu\to\bar{\nu}_\mu$ are neutrino oscillation channels through a long baseline. 
First, we perform the baseline and energy optimization of the neutrino
factory including the latest simulation results from the magnetized iron neutrino detector (MIND). 
Second, we study physics with near detectors and consider the treatment of systematic errors including cross section errors, flux errors, and background uncertainties.
Third, the effects of one additional massive sterile neutrino are investigated
in the context of near and far detector combinations.
\end{abstract}

\section{Introduction}
Neutrino oscillation experiments have provided compelling evidence
that the active neutrinos are massive particles~\cite{GonzalezGarcia:2007ib}, calling for physics beyond the Standard Model. Given three generations of massive neutrinos,
two mass-squared differences $\Delta m_{21}^2$ and $|\Delta m_{31}^2|$ are well understood while we have well-constrained mixing angles $\theta_{12}$, $\theta_{23}$ and $\theta_{13}$ so far~\cite{Schwetz:2011zk}, including the new results from T2K~\cite{Abe:2011sj} and MINOS ~\cite{Adamson:2011qu} hinting at a non-zero $\theta_{13}$. There are still unknown problems in the standard scenario: whether $\Delta m_{31}^2>0$ (normal ordering) or $\Delta m_{31}^2<0$ (inverted ordering); the value of $\theta_{13}$, and whether there is CP violation (CPV) in the lepton sector.

An exceptional LSND measurement has implied an $\mathcal{O}(1)$ eV$^2$ mass-squared difference~\cite{Aguilar:2001ty}, which naturally requires an additional sterile neutrino with $|\Delta m_{41}^2| \gg |\Delta m_{31}^2|$. A global fit to experimental data, however, is not in favor to this hypothesis~\cite{Maltoni:2007zf}. The recent results from MiniBooNE, however, are consistent with sterile neutrino oscillations in the antineutrino sector~\cite{AguilarArevalo:2010wv}. It becomes a crucial question whether the sterile neutrino exists or not.

A neutrino factory is able to answer questions in terms of three active neutrinos and find clues whether there are any sterile neutrinos or not. In the neutrino factory, electron neutrinos and muon neutrinos are produced by pure muon decays. Then signals are caused by wrong-charged muons arising from the chain $\mu^+\to\nu_e\to\nu_\mu\to\mu^-$ and the right-charged muons coming from the chain $\mu^+\to\bar{\nu}_\mu\to\bar{\nu}_\mu\to\mu^+$ (similar to $\mu^-$ chains), where $\nu_e\to\nu_\mu$ and $\bar{\nu}_\mu\to\bar{\nu}_\mu$ are neutrino oscillation channels through a long baseline. A magnetized detector, such as MIND, is proposed to fulfill the requirements of charge identifications. The baseline configurations in the International Design Study for the Neutrino Factory (IDS-NF)~\cite{ids} include the beam energy at $E_\mu=25 \, \mathrm{GeV}$ and two MINDs located at $L_1 \simeq 3 \,000 - 5
\, 000 \, \mathrm{km}$ and $L_2 \simeq 7 \, 500 \, \mathrm{km}$ (the
``magic'' baseline~\cite{Huber:2003ak}), respectively. In addition, there are no Near Detectors (NDs) in the IDS-NF.

\section{Update of optimization of the neutrino factory}
\begin{figure}[!tb]
\centering
\includegraphics[width=0.7\textwidth,height=0.6\textwidth]{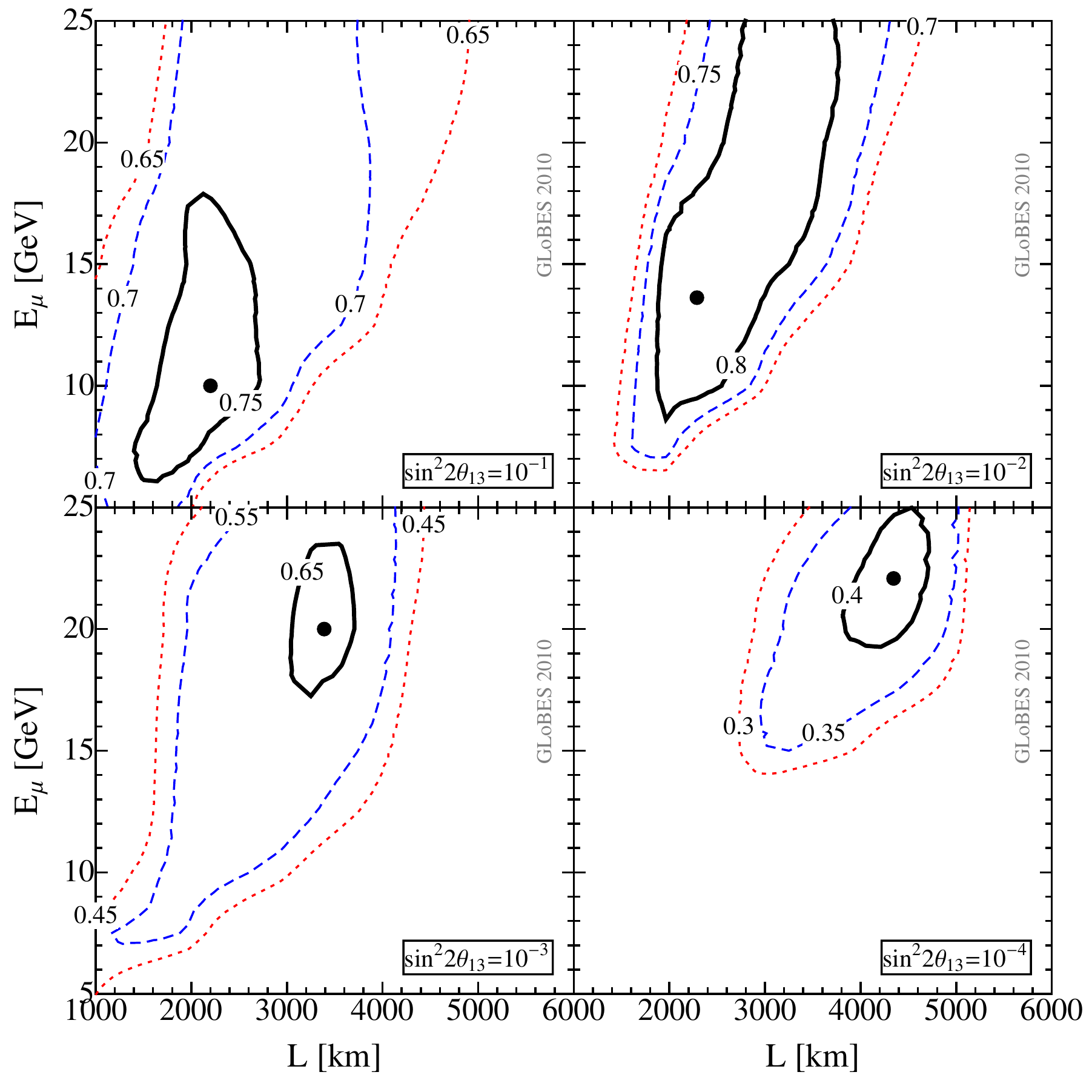}
\caption{\label{fig:LvsE}The discovery reach of CP violations (at $3 \sigma$ C.L.) as a function of the baseline $L$ and the beam energy $E_{\mu}$ for the single baseline neutrino factory with 50 kt detector. The contours are the fraction of CP violating $\delta_{\mathrm{CP}}$. The optimal performance is marked by a dot: $(2200,10.00)$, $(2288,13.62)$, $(3390,20.00)$ and $(4345,22.08)$ with regard to their best reaches of the fraction of $\delta_{\mathrm{CP}}$ at: 0.77, 0.84, 0.67 and 0.42. The figure is taken from the reference~\cite{Agarwalla:2010hk}.}
\end{figure}
 Recently more refined detector simulations have become available~\cite{Cervera:2010rz,ThesisLaing} and $\nu_e(\nu_\mu)\to\nu_\tau\to\tau\to\mu$ (similar to the other polarity) has been ignored for a long time~\cite{Indumathi:2009hg,Donini:2010xk}. There is a strong request for an re-optimization of the beam energy and baseline at the neutrino factory in light of new migration matrices for MINDs including the oscillated-$\nu_\tau$ related backgrounds.
We consider the so-called migration matrices bridging between the incident and reconstructed neutrino energies. It can also apply to tau related backgrounds. We update the optimization of the beam energy and baselines for the neutrino factory. From Fig.~\ref{fig:LvsE}, we can read that the low-energy and high-energy neutrino factory are two versions of the same experiment optimized for different parameter space. For large $\sin^22\theta_{13} \simeq 10^{-1}$, shorter baselines and lower energies are preferred. Compared to earlier analyses without background migrations, too high $E_\mu$ are in fact disfavored in the large $\sin^22\theta_{13}$ case.
As for a traditional high-energy neutrino factory with $E_\mu \simeq 20-25$~GeV, the baseline between $4 \, 000$ and $5 \, 000$~km is still preferred.
\section{Oscillation physics by active neutrinos with near detectors}
\begin{figure}[!tb]
\centering
\includegraphics[width=0.6\textwidth]{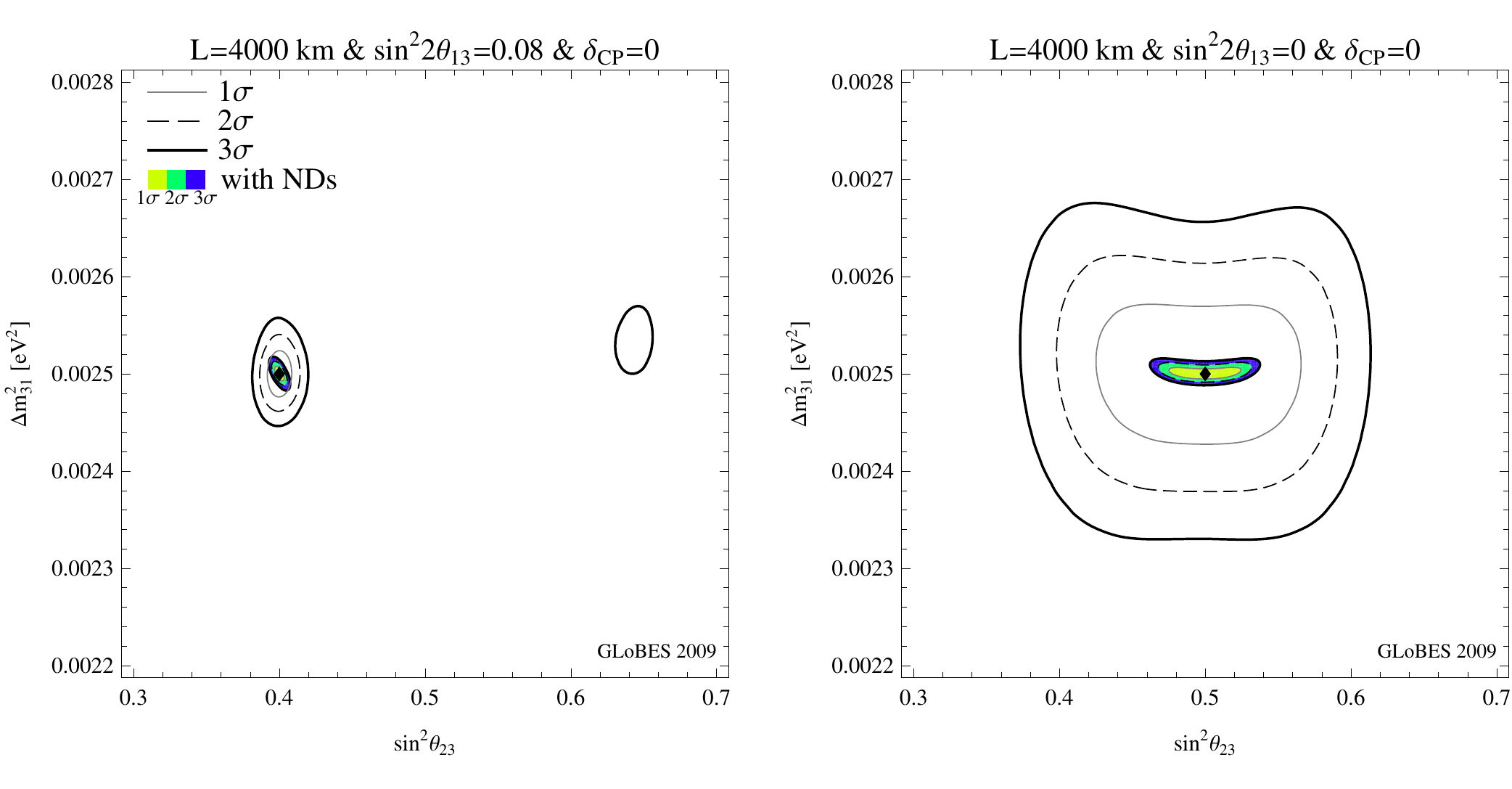}
\caption{\label{fig:nd}A comparison of the precision measurement at the $\sin^2\theta_{23}$--$\Delta m_{31}^2$ plane for the 4000 km baseline neutrino factory with or without NDs. The best fit points are marked by diamonds. The figures are adapted from \cite{Tang:2009na}.}
\end{figure}
The previous IDS-NF setup usually assumes the systematic errors coming from signal and background normalization errors uncorrelated among all channels and detectors. It is similar to the uncertainties for cross sections as well. We can take the $\nu_\mu\to\nu_\mu$ channel as an example to explain how near detectors cancel systematic uncertainties. We have the events for near detector (ND) and far detector (FD):
\begin{eqnarray}
 n_{\nu_\mu}^{\mathrm{ND}}=&\frac{N_{\mathrm{ND}}}{L_{\mathrm{ND}}^2}\Phi_{\nu_\mu}\sigma_{\nu_\mu}\epsilon_{\nu_\mu}\\
n_{\nu_\mu}^{\mathrm{FD}}=&\frac{N_{\mathrm{FD}}}{L_{\mathrm{FD}}^2}\Phi_{\nu_\mu}P(\nu_\mu\to\nu_\mu)\sigma_{\nu_\mu}\epsilon_{\nu_\mu}
\end{eqnarray}
We can immediately identify the dependence of $\sigma_{\nu_\mu}\epsilon_{\nu_\mu}$ cancels between near and far detectors after a combination of them:
\begin{equation}
 n_{\nu_\mu}^{\mathrm{FD}}=n_{\nu_e}^{\mathrm{ND}}\frac{N_{\mathrm{FD}}}{N_{\mathrm{ND}}}\frac{L_{\mathrm{ND}}^2}{L_{\mathrm{FD}}^2}P(\nu_\mu\to\nu_\mu)
\end{equation}
It implies that the uncertainties of cross sections and the detection efficiency will not take any effects assuming the same technology for ideal NDs and FDs. A refined treatment of systematic uncertainties including all channels is given in the reference~\cite{Tang:2009na}. As shown in Fig.~\ref{fig:nd}, we make a comparison of the precision measurement in the $\sin^2\theta_{23}$--$\Delta m_{31}^2$ plane for the 4000 km baseline neutrino factory with or without NDs. We find a significant improvement of sensitivity once we include NDs.

\section{Combine near and far detectors to search for sterile neutrinos}
\begin{figure}[!tb]
\centering
\includegraphics[width=0.8\textwidth]{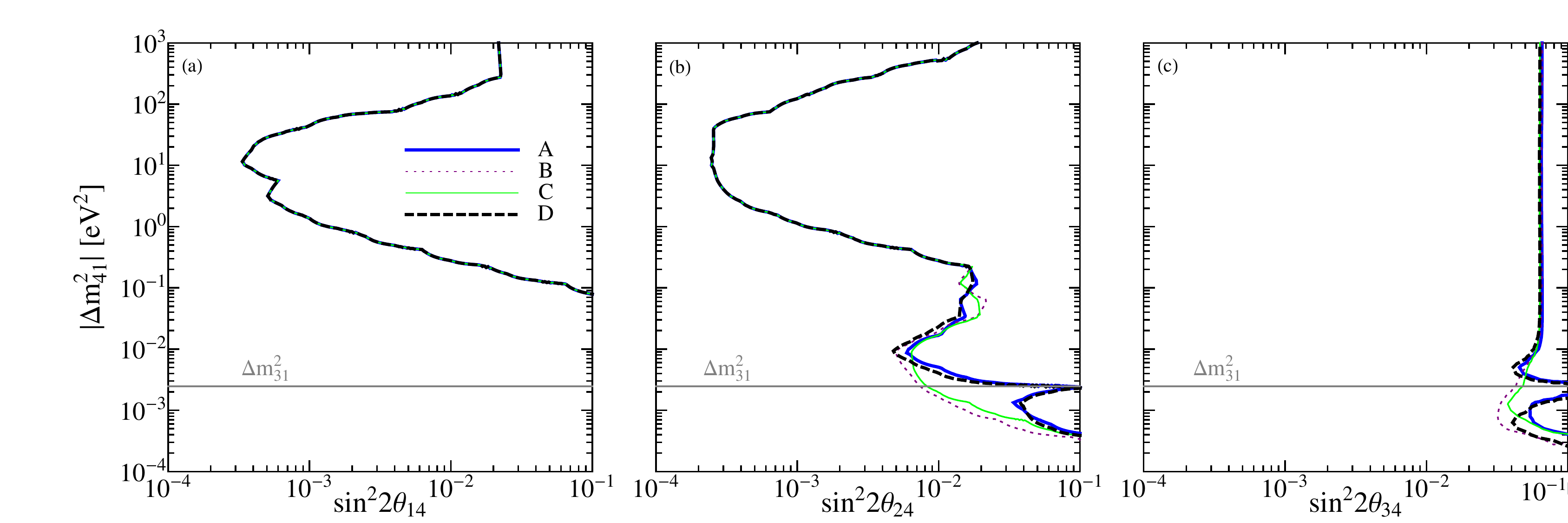}
\caption{\label{fig:sterile}The exclusion limit of $\sin^22\theta_{i4}$--$\Delta m_{41}^2\, (i=1,2,3)$ in terms of arbitrarily massive sterile neutrinos. Four different mass orderings are taken into account such as $\ldm>0$, $\vldm>0$ (A), $\ldm>0$, $\vldm<0$ (B), $\ldm<0$, $\vldm>0$ (C), and $\ldm<0$, $\vldm<0$ (D). The figure is taken from \cite{Meloni:2010zr}.}
\end{figure}
While earlier studies have focused on sterile neutrinos with a LSND-like mass splitting $\mathcal{O}(\mathrm{eV}^2)$ (see the references~\cite{GonzalezGarcia:2007ib,Maltoni:2007zf,Maltoni:2001bc}), we cannot rely on constraints from the atmospheric and solar experiments, which need to be re-analyzed in a global fit for the presence of very light sterile neutrinos in a self-consistent way. Therefore, we combine both near and far detectors to search for arbitrarily massive sterile neutrinos in a self-consistent treatment. In terms of one additional sterile neutrino, there are four different scenarios corresponding to $\ldm>0$, $\vldm>0$ (A), $\ldm>0$, $\vldm<0$ (B), $\ldm<0$, $\vldm>0$ (C), and $\ldm<0$, $\vldm<0$ (D). They are shown in the caption of Fig.~\ref{fig:sterile}. We consider the following configurations: two near detectors -- each with a mass of $32 \, \mathrm{t}$,
and each with a distance of $d=2 \, \mathrm{km}$ from the end of the straight section in the muon storage ring, combined with two far MINDs with one at 4000 km and the other at 7500 km. As shown in Fig.~\ref{fig:sterile}, the upper peak hardly depends on the mass ordering. The lower (long-baseline) peak, which is only present in the middle and right panels, depends on the mass ordering. There are two qualitatively different cases: for the schemes A and D, the sensitivity is destroyed just at the value of $\ldm$ while the exclusion limits for the schemes B and C remain unchanged.

\section{Conclusions}
The neutrino factory is one of the most powerful machine towards neutrino oscillation physics. We have revisited the optimization of the beam energy and the baselines in light of the latest information. We have found that the low-energy and high-energy neutrino factory are just two versions of the same experiment optimized for a different parameter space. In addition, we have discussed how to adopt NDs to cancel systematic uncertainties. We have shown the impact on NDs at the precision measurement of $\theta_{23}$ and $\Delta m_{31}^2$. Finally, we have investigated the search for sterile neutrinos by combining NDs and FDs at the neutrino factory. A substantial difference of four different mass orderings has been presented.
\section*{Acknowledgements}
JT appreciates Walter Winter's collaborations and help during his Ph.D period. JT is grateful to Prof. Lothar Oberauer for the invitation and good organization of TAUP2011.
JT was supported to attend the conference by DFG Research Training Group GRK1147 and DPG WE Heraeus grant.

\section*{References}
\bibliographystyle{iopart-num}
\bibliography{references}

\providecommand{\newblock}{}
\begin{thebibliography}{10}
\expandafter\ifx\csname url\endcsname\relax
  \def\url#1{{\tt #1}}\fi
\expandafter\ifx\csname urlprefix\endcsname\relax\def\urlprefix{URL }\fi
\providecommand{\eprint}[2][]{\url{#2}}

\bibitem{GonzalezGarcia:2007ib}
Gonzalez-Garcia M~C and Maltoni M 2008 {\em Phys. Rept.\/} {\bf 460} 1--129
  (\textit{Preprint} \eprint{0704.1800})

\bibitem{Schwetz:2011zk}
Schwetz T, T\'ortola M and Valle J 2011 {\em New J. Phys.\/} {\bf 13} 109401
  (\textit{Preprint} \eprint{1108.1376})

\bibitem{Abe:2011sj}
Abe K {\em et~al.\/} (T2K Collaboration) 2011 {\em Phys. Rev. Lett.\/} {\bf
  107} 041801 (\textit{Preprint} \eprint{1106.2822})

\bibitem{Adamson:2011qu}
Adamson P {\em et~al.\/} (MINOS Collaboration) 2011 {\em Phys. Rev. Lett.\/}
  {\bf 107} 181802 (\textit{Preprint} \eprint{1108.0015})

\bibitem{Aguilar:2001ty}
Aguilar A {\em et~al.\/} (LSND Collaboration) 2001 {\em Phys. Rev.\/} {\bf D64}
  112007 (\textit{Preprint} \eprint{hep-ex/0104049})

\bibitem{Maltoni:2007zf}
Maltoni M and Schwetz T 2007 {\em Phys. Rev.\/} {\bf D76} 093005
  (\textit{Preprint} \eprint{0705.0107})

\bibitem{AguilarArevalo:2010wv}
Aguilar-Arevalo A~A {\em et~al.\/} (MiniBooNE Collaboration) 2010 {\em Phys.
  Rev. Lett.\/} {\bf 105} 181801 (\textit{Preprint} \eprint{1007.1150})

\bibitem{ids}
 International design study of the neutrino factory {\tt http://www.ids-nf.org}

\bibitem{Huber:2003ak}
Huber P and Winter W 2003 {\em Phys. Rev.\/} {\bf D68} 037301
  (\textit{Preprint} \eprint{hep-ph/0301257})

\bibitem{Agarwalla:2010hk}
Agarwalla S~K, Huber P, Tang J and Winter W 2011 {\em JHEP\/} {\bf 01} 120
  (\textit{Preprint} \eprint{1012.1872})

\bibitem{Cervera:2010rz}
Cervera A, Laing A, Mart{\'i}n-Albo J and Soler F~J~P 2010 {\em Nucl. Instrum.
  Meth.\/} {\bf A624} 601--614 (\textit{Preprint} \eprint{1004.0358})

\bibitem{ThesisLaing}
Laing A 2010 {\em Optimization of Detectors for the Golden Channel at a
  Neutrino Factory\/} Ph.D. thesis Glasgow University

\bibitem{Indumathi:2009hg}
Indumathi D and Sinha N 2009 {\em Phys. Rev.\/} {\bf D80} 113012
  (\textit{Preprint} \eprint{0910.2020})

\bibitem{Donini:2010xk}
Donini A, Gomez~Cadenas J~J and Meloni D 2011 {\em JHEP\/} {\bf 1102} 095
  (\textit{Preprint} \eprint{1005.2275})

\bibitem{Tang:2009na}
Tang J and Winter W 2009 {\em Phys. Rev.\/} {\bf D80} 053001 (\textit{Preprint}
  \eprint{0903.3039})

\bibitem{Meloni:2010zr}
Meloni D, Tang J and Winter W 2010 {\em Phys. Rev.\/} {\bf D82} 093008
  (\textit{Preprint} \eprint{1007.2419})

\bibitem{Maltoni:2001bc}
Maltoni M, Schwetz T and Valle J~W~F 2002 {\em Phys. Rev.\/} {\bf D65} 093004
  (\textit{Preprint} \eprint{hep-ph/0112103})

\end{thebibliography}

\end{document}